\documentclass[sigconf,
    backend=biber,
    natbib=false,            
    compact]{acmart}
\usepackage[url=false,             
    doi=false, 
    isbn=false]{biblatex} 
\renewcommand\footnotetextcopyrightpermission[1]{}
\usepackage{subcaption} 
\usepackage{amsmath}
\usepackage{float}
\usepackage{subcaption}
\usepackage{booktabs}
\usepackage{graphicx}

\begin{document}
\title{Selecting User Histories to Generate LLM Users for Cold-Start Item Recommendation}
\pagestyle{plain} 

\author{Nachiket Subbaraman}
\email{nsubbaraman@ucdavis.edu}
\affiliation{%
  \institution{University of California, Davis}
  \city{Davis}
  \state{CA}
  \country{USA}
}

\author{Jaskinder Sarai}
\email{jssarai@ucdavis.edu}
\affiliation{%
  \institution{University of California, Davis}
  \city{Davis}
  \state{CA}
  \country{USA}
}

\author{Aniruddh Nath}
\email{aniruddhnath@google.com}
\affiliation{%
  \institution{Google Inc.}
  \city{Mountain View}
  \state{CA}
  \country{USA}
}

\author{Lichan Hong}
\email{lichan@google.com}
\affiliation{%
  \institution{Google DeepMind}
  \city{Mountain View}
  \state{CA}
  \country{USA}
}

\author{Lukasz Heldt}
\email{heldt@google.com}
\affiliation{%
  \institution{Google Inc.}
  \city{Mountain View}
  \state{CA}
  \country{USA}
}

\author{Li Wei}
\email{liwei@google.com}
\affiliation{%
  \institution{Google Inc.}
  \city{Mountain View}
  \state{CA}
  \country{USA}
}

\author{Zhe Zhao}
\email{zao@ucdavis.edu}
\affiliation{%
  \institution{University of California, Davis}
  \city{Davis}
  \state{CA}
  \country{USA}
}

\begin{abstract}
Large Language Models (LLMs) have demonstrated remarkable capabilities in reasoning, generalization, and simulating human-like behavior across a wide range of tasks. These strengths present new opportunities to enhance traditional recommendation systems (RS), especially in the cold-start item scenario where newly introduced items lack interactions. Existing works have used LLMs to address cold-start issues in traditional RS through data augmentation, but they have limitations. One recent work directly addresses this issue by prompting LLMs to generate augmented interaction data between randomly sampled users and cold-start items. Then, they train the traditional RS with augmented data, incorporating collaborative signals for cold-start items. Although they use LLMs to provide cold-start items with feedback, they use partial user histories, which does not allow the LLM to fully emulate the user. Furthermore, randomly selecting users  is not optimal for augmentation. To address these challenges, we leverage the LLM as a user and develop a reinforcement learning (RL) framework that trains a policy to select users for augmentation, optimizing for cold-start item performance after augmented training. The policy model learns to select users for cold-start item data augmentation based on their behavioral features and histories. To optimize user selection for cold-start item performance, we employ a policy gradient method that updates the policy in the direction of actions that lead to high rewards. Experiments on Amazon Product Review  datasets show substantial gains in cold-start item recall, demonstrating the effectiveness of our method as a scalable, serving-efficient augmentation strategy for modern RS.
\end{abstract}

\maketitle
\section{Introduction}
RS are ubiquitous in modern digital platforms, driving user engagement by serving interesting content tailored to user preferences. Traditionally, collaborative filtering (CF) has been used to recommend items using communal user feedback \cite{Schafer2007}. Most RS models rely on well-trained ID-embeddings to learn and predict user-item relevance \cite{yi2019sampling, he2017neuralcollaborativefiltering}. However, the ID-embedding models are unable to provide high-quality recommendations for cold-start scenarios \cite{zhao2022improving}. There are three categories of the cold-start recommendation problem: a) new users, b) new items, or c) new users and new items \cite{LIKA20142065}. Though cold-start users is a well-known challenge, the cold-start item problem is particularly critical. Millions of new items are uploaded to recommendation platforms daily \cite{10.1145/3041021.3054202, chen2024multi, davidson2010youtube}. Their lack of interaction data makes it difficult for CF models to learn their embeddings and recommend them to users. 

To improve cold-start item recommendations, past studies have employed item content features or ID embedding dropout \cite{10.1145/3580305.3599519, li2023gpt4recgenerativeframeworkpersonalized, WEI201729, 10.1145/3124791.3124792, li2022surveydropoutmethodsexperimental, NIPS2017_dbd22ba3}. For example, several methods randomly assign cold-start item IDs to hash buckets during training and initialize them with their corresponding hash buckets during serving time \cite{yi2019sampling, kang2021learning, shiao2024improvingoutofvocabularyhandlingrecommendation}. This approach is widely used in the industry to avoid serving randomly initialized embeddings for cold-start IDs. 
Recently, LLMs have introduced human-like reasoning and natural language generation capabilities. Consequently, a new line of work has applied LLMs to RS, including the cold-start problem. One work fine-tunes LLMs to simulate cold-start item recommendations \cite{huang2025large}. Another work directly uses LLMs to recommend items in a cold-start setting \cite{sanner2023large}. However, fine-tuning and serving LLM requests for cold-start items are time-consuming and expensive \cite{lin2024data}. Further, unlike traditional RS, LLM-based recommenders have difficulty capturing collaborative information, even after fine-tuning on the recommendation data \cite{kim2024large, bao2024large}.

An alternative approach to the cold-start item problem leverages LLM as a data augmenter for a traditional RS backbone \cite{wang2024largelanguagemodelsdata, 10.1145/3640457.3688159}. To generate collaborative information for cold-start items, \cite{wang2024largelanguagemodelsdata} uses LLM to generate cold-start item interactions, then trains RS models with the augmented data. This approach avoids the LLM training complexity and inferior CF issues from which the LLM-as-RS paradigm suffers. However, they randomly sample (user, item) interactions and prompt the LLM using only the items that the user interacted with before that interaction, thereby excluding any items the user engaged with afterward. Another similar work prompts the LLM to provide the types of items that the user prefers given the m highest-rated items from the user's interaction history, which omits the items that the user rated poorly \cite{10.1145/3640457.3688159}. 
\begin{figure*}[!t]
\includegraphics[width=0.8\linewidth,height=6.5cm]{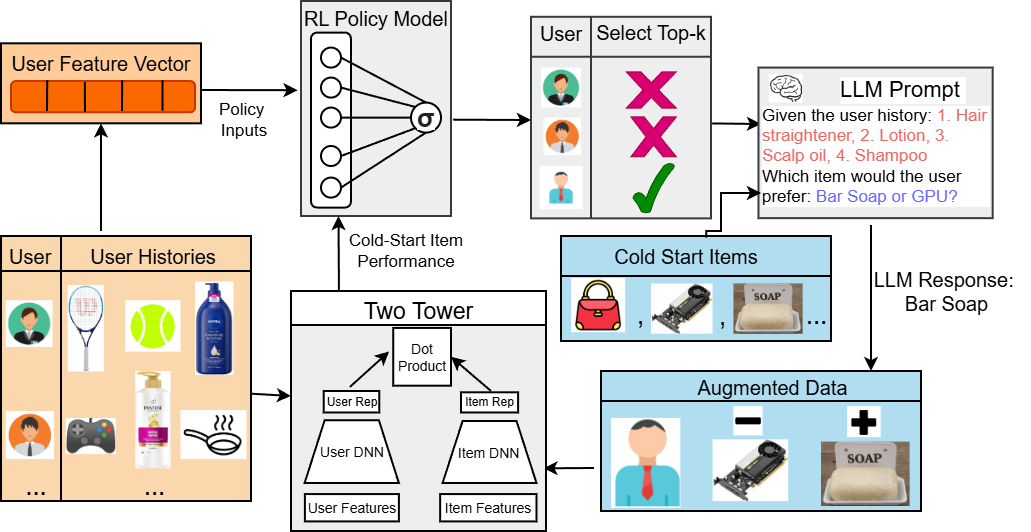}
    \caption{Overview of our method which leverages a policy model to select users for LLM-based data augmentation. We provide user features to a policy and train it to maximize cold-start item performance using a novel policy gradient method.}
    \label{fig:overall_framework}
\end{figure*}
Since both prompts omit historical item interactions that would improve the LLM's ability to imitate the user, we need to address the question: "\textit{How do we generate augmentation samples that reflect the user’s true preference}"? To enable the LLM to act as the user, we sample users instead of interactions and provide the LLM with the user's entire interaction history.

Further, Wang et al. \cite{wang2024largelanguagemodelsdata} prompts the LLM to predict which of two randomly sampled cold-start items a user would prefer given a randomly sampled interaction’s history. Randomly sampling users for cold-start item data augmentation could result in widely varying cold-start item performances depending on the characteristics of the selected users. Thus, selecting the right users for such augmentation is an open challenge and raises the question: “\textit{Which users most improve cold-start item performance when selected for data augmentation}”? To answer this question, we propose an optimization framework that selects users for cold-start item data augmentation that improve cold-start item performance. In this work, we introduce a scalable LLM-based data augmentation framework that trains a policy model to maximize the probability of selecting users that generate high rewards, similar to REINFORCE \cite{10.5555/3009657.3009806}. We illustrate our overall framework in Figure \ref{fig:overall_framework}.

We conduct experiments on two distinct, real-world datasets to demonstrate the superiority of our RL policy for cold-start item recommendations. First, we evaluate cold-start item recall without any data augmentation. Then, we select random users and users with the highest values per user feature for augmentation. Lastly, we select users for augmentation using our RL policy, showing an improvement over the prior selection strategies. Additionally, we split the test set into examples with selected and unselected users and evaluate cold-start item performance on both partitions. Our contributions are summarized as follows: 
\begin{enumerate}
\item We enable the LLM to impersonate users, providing more realistic feedback for cold-start items. 
\item We introduce a policy gradient method that improves cold-start item performance by selecting optimal users for LLM-based data augmentation. 
\item We perform extensive experiments demonstrating that the learned policy outperforms other user selection methods for cold-start item data augmentation and evaluate the effects of augmentation on selected and unselected users. 
\item We expand our policy model and improve training efficiency via proxy rewards.
\end{enumerate}
We designed our method to address the cold-start item problem effectively and efficiently. In Section \ref{sec:related-works}, we provide a literature review addressing the cold-start item problem, integrating RS with LLMs, and employing RL techniques for RS. We provide an overview of RS and policy gradients in Section \ref{sec:preliminaries}. In Section \ref{sec:our-method}, we explain our overall framework, including the policy model, reward formulation, objective computation, and gradient update. For our RS backbone, we use the widely adopted two-tower model \cite{yi2019sampling}. We present our results in Section \ref{sec:experiments} and conclude our study in Section \ref{sec:conclusion}. 

\section{Related Works}
\label{sec:related-works}
\subsection{Cold-Start Item Problem}
Cold-start items lack collaborative behavior data to train high-quality ID embeddings. Existing works have proposed generative models that approximate the behavior embedding of cold-start user/item IDs given their content embedding, which closes the gap between warm and cold ID embeddings \cite{10.1145/3626772.3657839,NIPS2013_b3ba8f1b,chen2022generative,10.1145/3539618.3591732}. Recently, Semantic IDs have outperformed randomly hashed IDs by generating dense item representations through pre-trained encoders applied to item content embeddings, resulting in improved generalization abilities on cold-start items \cite{singh2024better, rajput2023recommender, penha2025semantic, zheng2025enhancing}. However, none of these approaches directly address the lack of cold-start item interactions. 

Unlike traditional recommender systems that rely on historical user-item interactions, recent approaches rely on LLMs to generate cold-start item embeddings. For example, ColdLLM \cite{10.1145/3701551.3703546} uses LLMs to simulate the historical behaviors of cold-start items. Also, FilterLLM \cite{liu2025filterllm} uses the next-token capabilities of LLMs to predict user interaction distribution for cold-start items. However, these methods require fine-tuning massive foundation models, which incurs higher computational burden than traditional ID-based recommenders. Further, they decouple the generated cold embeddings from the latent collaborative information present in warm embeddings, which can negatively impact recommender performance. Our method utilizes less-expensive LLM inference requests to generate augmented data and train a traditional RS model. 

\subsection{Recommendation with LLMs}
Recent advances in LLMs have introduced new paradigms for recommendation, merging traditional collaborative filtering models with LLMs \cite{luo2024integrating,10.1007/978-3-031-56060-6_24}. LLMs can act as agents capable of human-like behavior, including chatting, reasoning, and in-context learning \cite{kong2023better}. These abilities have unlocked rich natural language data sources within recommendation systems, including item descriptions and reviews, which are more expressive than nontextual user-item interactions \cite{deldjoo2024recommendation}. Since collaborative filtering cannot capture user interests for cold-start items, we can leverage LLMs to infer user preferences using natural-language based personalization.

Existing LLM for recommendation works prompt LLMs to zero-shot rank candidate items and encode user/item features into dense embeddings for downstream RS models \cite{10.1007/978-3-031-56060-6_24}. Other studies generated end-to-end recommendations directly without any external models \cite{sanner2023large}. Another line of works rely on LLM-based agents to converse with users or simulate their behaviors \cite{zhang2024generative,friedman2023leveraging,wang2025user,10.1145/3589334.3645537}. For example, a study similar to ColdLLM simulates cold-start users by generating their histories given their demographic data \cite{10.1145/3705328.3748109}.
However, there are massive computational costs associated with LLM requests in large-scale RS. The number of LLM inference requests we perform is a fraction of the number of users. This reduces expenses and improves computational efficiency in production environments. Furthermore, LLMs are pretrained on a broad corpora where frequently mentioned items dominate, causing them to overemphasize popular content and disadvantage less-represented items in RS \cite{chen2021biasdebiasrecommendersystem}. In contrast, our approach leverages LLMs to simulate user preference between only two randomly selected cold-start items, mitigating popularity bias. While LLMs show promise in zero-shot settings, conversational RS, and user behavior simulation, their effectiveness in cold-start scenarios remains underexplored. 

\subsection{RL for Recommendations}
Existing works have implemented reinforcement learning (RL)-based approaches to simulate interactions in recommendation environments \cite{ie2019recsimconfigurablesimulationplatform,DBLP:journals/corr/abs-1808-00720}. Other works applied RL techniques to optimize RS, similar to our contribution. For example, one study scales the REINFORCE policy gradient to a production-grade RS and addresses biases by training an off-policy correction \cite{10.1145/3289600.3290999}. A similar work optimizes long-term user engagement using a model-based RL framework \cite{10.1145/3292500.3330668}. RL methods that address the cold-start item problem optimize item lifetime value \cite{10.1145/3459637.3482292} and simulate user interactions to optimize cold-start item embeddings \cite{liu2024fine}.

However, their RL agents operate on a large action space, typically including all items, which introduces significant complexity in large-scale recommendation systems. On the other hand, our action space is $\{0, 1\}$, which is more manageable. Further, the existing works translate recommendation into a Markov Decision Process (MDP) setup, where the decision to enter a new state depends on the current state and action. Instead, we used a contextual-bandits setup for our method, because we select each user independently, using only their features. Moreover, policy-gradients have been successfully applied to a contextual-bandits setup in RS, which justifies our decision \cite{10.1145/3308558.3313616}.

\section{Preliminaries}
\label{sec:preliminaries}
\textbf{Recommendation System.} In this work, we consider the retrieval task operating under implicit feedback. Let $U$ and $I$ denote the set of users and items. The observed interaction matrix $R \in \{0,1\}^{|U|x|I|}$ encodes binary feedback: $r_{u,i} = 1$ if user u interacts with item i and $r_{u,i} = 0$ otherwise. This setup reflects a real-world scenario with clicks, views, or purchases but no explicit ratings. Each user and item is assigned an ID embedding $e_u, e_i \in R^m$, indexed via a one-hot encoding into user and item embedding tables. These embeddings are trained via matrix factorization to predict the observed interactions with the dot product of their corresponding latent factor vectors. For example, if the latent factor vectors corresponding to user $u$ and item $i$ are $l_u$ and $l_i$, then the model predicts their compatibility via $\hat{r}_{u,i} = l_u^\mathrm{T}l_i$. In cold-start settings, items lack sufficient interaction history, making it difficult to learn their embeddings.\\
\textbf{Policy-Gradient.}
To address the cold-start item problem, we formulate user selection as a policy-gradient problem, which enables the optimization of cold-start item performance through backpropagation. Unlike value-based approaches, policy gradient methods converge to locally optimal policies \cite{10.5555/3009657.3009806}. Here is the general objective function formula: 
\begin{equation}
J(\theta) = \mathbb{E}_{\tau \sim{\pi_\theta}} [R(\tau)]
\end{equation}
The expectation is taken over a trajectory $\tau$ of state, action pairs $(s_0, a_0),...,(s_k, t_k)$ sampled from a policy $\pi_\theta$ parametrized by $\theta \in \mathbb{R}^d$. Thanks to the "log-derivative" trick, we can derive the gradient of the objective with respect to $\theta$:
\begin{equation}
\nabla_{\theta}J(\theta) = \mathbb{E}_{\tau \sim{\pi_\theta}}[R(\tau)\nabla \log(\pi_\theta(\tau))]
\label{eq:policy-gradient}
\end{equation}
Equation \ref{eq:policy-gradient} yields the common REINFORCE method \cite{williams1992reinforce}, which stochastically picks actions and increases the probability of selecting high-return states.

\section{Our Method}
\label{sec:our-method}
Our method uses LLMs to generate synthetic interactions between selected users and cold-start items. First, we pass the user features into the policy model. Then, the policy model processes features through a single linear layer and sigmoid activation to compute scores per-user that estimate their suitability for data augmentation. We iterate over the users in descending score order and select each user based on their sigmoid-derived selection probabilities. For each sampled user, we prompt the LLM to produce cold-start item pairwise preferences, generating the augmented data. Finally, we train two-tower models with the original and augmented data, which generates cold-start item performance rewards. The policy gradient uses the rewards to update the parameters in a direction that increases the probability of selecting high-reward users. These steps are executed iteratively until policy convergence. We describe each component in the following sections. 

\subsection{Prompt LLM to Act As a User}
As mentioned in the introduction, we prompt the LLM with the user’s entire training history, unlike previous LLM-based data augmentation works \cite{wang2024largelanguagemodelsdata}. This enables the LLM to generate interactions in a manner that is more faithful to the user's persona. Using only the interaction history biases the LLM towards specific items that the user reviewed, ignoring the items that the user reviewed after the interaction timestamp. This would cause the LLM persona to be less faithful to the user and make inaccurate cold-start item preference predictions. 

\subsubsection{Prompt Construction}
\begin{figure}[htb]
    \includegraphics[width=8cm]{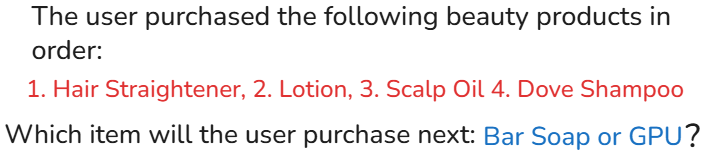}
    \caption{Example LLM User Prompt}
    \label{fig:sample_LLM_prompt}
\end{figure}

Adapted from a prior work \cite{wang2024largelanguagemodelsdata}, we provide the LLM with a pair of randomly sampled cold-start items to generate pairwise preference augmentation examples. Each item in the user's history is represented by their title in the LLM prompt. For example, Figure \ref{fig:sample_LLM_prompt} illustrates the cold-start item titles with the blue highlighted text, while the red highlighted item titles represent the user’s history.
When computing the augmentation loss (explained in Section \ref{sec:train-recsys}), the item that the LLM picks is considered the positive cold-start item while the other item is considered the negative.

\subsection{Policy Model Framework}
\begin{figure}[t]
    \includegraphics[width=\linewidth]{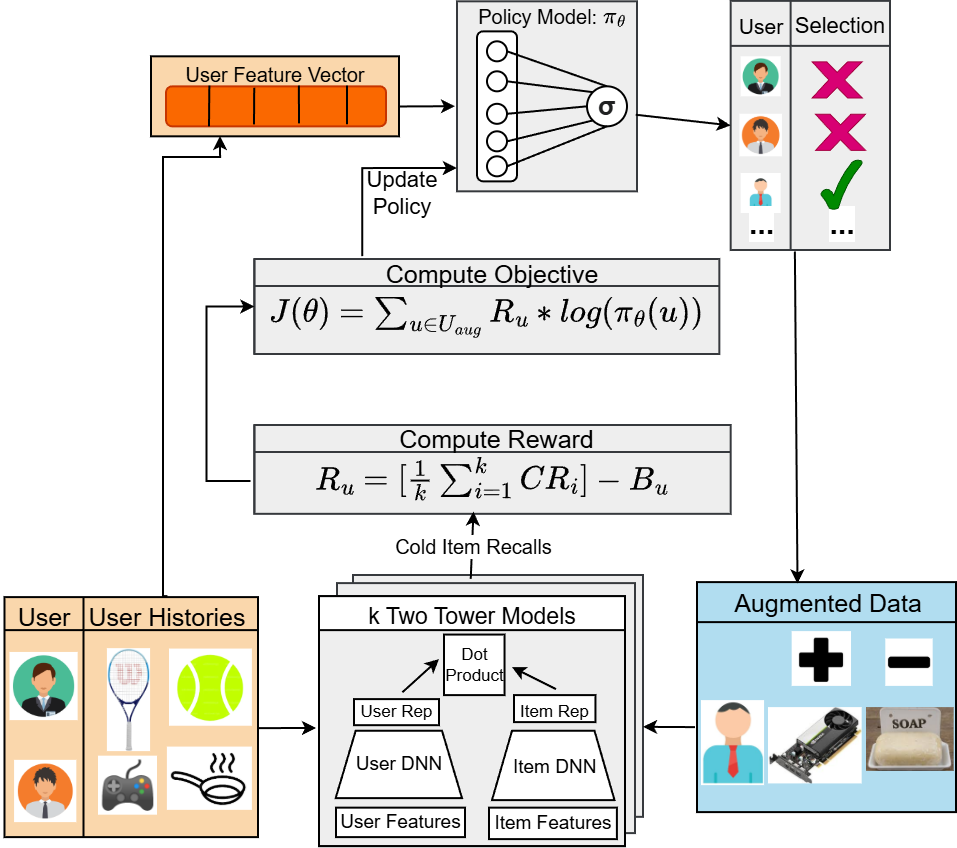}
    \caption{Policy Model Training Framework}
    \label{fig:policy_model}
\end{figure}
To address the cold-start item problem, we propose a policy model which selects users for cold-start item data augmentation. We formulate the user selection process within a reinforcement learning paradigm, as illustrated in Figure \ref{fig:policy_model}.\\
\textbf{Context Space.} At each time step, the policy observes a user's features represented by a feature vector $f_u \in \mathbb{R}^d$, where $d$ is the feature vector dimension. \\
\textbf{Action Space.} Given the context $f_u$, the policy decides whether or not to select user $u$ for augmentation. Thus, the action space is $\mathcal{A} =\{0, 1\}$, where $0$ and $1$ represent unselected and selected, respectively. \\
\textbf{Policy Model.} Our policy model, $\pi_\theta$, is parametrized by $\theta \in \mathbb{R}^d$ and operates over $U$. It computes the probability of selecting a user for augmentation. For a given user $u \in U$, $\pi_\theta(a=1|f_u) = \sigma(\frac{f_u \theta^\top}{T})$. Here, $\sigma$ represents the sigmoid function. While this formula is derived under a linear policy assumption, we also test a two-layer policy and report both sets of results in Section \ref{sec:experiments}. We tuned the temperature $T$ to control the exploration-exploitation tradeoff. 

We iterate through the users $u \in U$ in descending order of their output logits, $f_u\theta ^\top$. For each user, we select them with probability $\pi_\theta(a=1|f_u)$. Once we select a user subset of specified size, we generate our augmentation data by executing LLM inference requests on each of the selected users. Then, we train two-tower models to compute our policy reward, objective, and update the parameters via gradient ascent. 

\subsubsection{User Features}
We created user features, using users' historical item metadata, to capture distinct user characteristics. For each user feature, we first select the users with the highest feature values (top-k users) for data augmentation. Then, we train two-tower models with the original and augmented training data. The training results enable us to examine the characteristics of users that improve cold-start item performance when selected for data augmentation. The features whose top-k users yield the highest cold recalls when selected for augmentation are min-max scaled to $[0,1]$ and provided to the policy model as inputs. We describe the features that we provide to the policy in Table \ref{tab:user-feature-definitions} and justify their usage in Appendix \ref{sec:user-features-rationale}. For each feature, we select its top-k users for cold-start item augmentation and evaluate their performance in Section \ref{sec:experiments}. 

\subsubsection{Reward Generation}
\label{sec:reward_generation}
To generate distinct rewards per user, we incorporate a baseline, which is a standard practice to reduce variance in the REINFORCE gradient estimate \cite{10.5555/3009657.3009806}. First, we compute each user's initial reward baseline as the average cold-start item recall@K over $M$ non-augmented two-tower models. 
\begin{equation}
B_0 = \frac{1}{M}\sum_{i=1}^{M} CR@K^{(i)}
\label{eq:avg-cr}
\end{equation} 
Cold-start item recall@K computes recall@K only on the test set examples where the positive item is cold. We refer to this metric as cold recall (or CR). Reward baseline $B_0$ ensures that early updates reflect meaningful improvements. 
To refine this baseline initialization, we introduce the average cold recall@K generated by selecting the top-k users per user feature for augmentation. 
\begin{table*}[!htp]
\centering
\caption{Definitions and abbreviations of user features passed into the policy.}
\begin{tabular}{llp{9cm}}
\toprule
\textbf{Feature} & \textbf{Abbreviation} & \textbf{Definition} \\
\midrule
Median Popularity & MP & Median number of reviews for each item the user reviewed \\
Average Rating & AR & Average ratings of the user's reviews\\
Average Popularity & AP & Average number of reviews for each item the user reviewed \\
Rating Variance & RV & Variance of the user's ratings \\
Category KL Diversity & CKLD & KL-divergence between the user’s historical and global item categories \\
Category Simpson Diversity & CSD & Sum of the squares of user’s historical item category probabilities \\
Embedding Entropy & EE & Vendi score \cite{friedman2023vendiscorediversityevaluation} of a user’s historical item metadata BERT embeddings \\
Velocity & V & Number of users that reviewed an item soon after the user reviewed it \\
\bottomrule
\end{tabular}
\label{tab:user-feature-definitions}
\end{table*}
For example, let $U_{RV}$ represent the set of users with the top-k RV feature values (refer to Table \ref{tab:user-feature-definitions} for user feature names). Then, we use $U_{RV}$ to generate cold-start item augmentation data, train $M$ two-tower models, and compute the average cold recall@K, $CR_{RV}$, as mentioned in Equation \ref{eq:avg-cr}. Once we have the average cold recalls per feature selection, we initialize each user's baseline as the average cold recall across their features. For example, if user $u\in U_{RV} \cap U_{AP} \cap U_{MP}$, then the feature-averaged recall for $u$ is 
\begin{equation}
F_u = \frac{1}{3} [[CR@K]_{RV} + [CR@K]_{AP} + [CR@K]_{MP}]
\end{equation} 

Importantly, we provide the policy with the user features yielding the highest $CR@K$s. Also, this aggregation is restricted to the subset of features used as inputs to the policy model. Including features outside the policy space would introduce noise and potentially bias the reward signals towards extraneous dimensions. 
Finally, we refine each user's baseline by computing an exponential moving average (EMA) between the non-augmented cold recalls and feature-averaged cold recalls. 
\begin{equation}
B_{u} = \alpha B_{0} + (1-\alpha) F_{u}, 0 \leq \alpha \leq 1, \forall u \in U
\end{equation}
This approach avoids the additional parameters and overfitting risk of training a critic baseline, while assigning distinct credit to each user. To smoothen the reward signal and enforce an upward trajectory of cold recalls, we also use the EMA to update each user's baseline during policy training. For each policy iteration, equation \ref{eq:reward-baseline} computes the reward for user $u$ by subtracting their baseline from the average cold recall@K across $M$ two-tower models. 
\begin{equation}
\label{eq:reward-baseline}
R_u = [\frac{1}{k} \sum_{i=1}^{M} CR@K^{(i)}] - B_u
\end{equation}
\subsubsection{Objective Computation}
We formulate the policy objective according to the REINFORCE policy gradient with a few simplifications. Namely, we do not use any discount factor or trajectories. This formulation aligns with a contextual bandits setup, where each user’s selection decision is independent and conditioned on user features, rather than a sequential decision process modeled by a MDP. First, we compute the objective using the per-user rewards and their selection probabilities: \\
\begin{equation}
J(\theta) = \sum_{i=1}^{|U_s|} R(u_i) \times \log(\pi_{\theta}(a=1|f_{u_i})) \\
\end{equation}
Here, $U_s$ is the set of selected users. Since the reward signals $R(u_i)$ are constants with respect to $\theta$, the policy gradients are computed using the following formula:
\begin{equation}
\nabla_{\theta} J({\theta}) = \sum_{i=1}^{|U_s|} R(u_i) \times \nabla_{\theta}\log(\pi_{\theta}(a=1|f_{u_i})) 
\end{equation}\\
Then, we update the policy parameters with learning rate $\eta$:
\begin{equation}
\theta_{t+1} = \theta_{t} + \eta \times \nabla_{\theta} J(\theta)
\end{equation}
\subsection{Train RS with Augmentation}
\label{sec:train-recsys}
As a baseline for our experiments, we used the Two-Tower retrieval model \cite{yi2019sampling}. The user tower’s input layer only encodes the user ID embedding, while item tower’s input layer concatenates the item ID embedding with the item metadata embedding. We use BERT to encode item titles, descriptions, brands, categories, and features \cite{hou2022towards}. To train the Two-Tower model, we use the In-Batch Softmax (IBS) with LogQ correction \cite{yi2019sampling} instead of the full softmax to improve training speed. To implement the LogQ correction, we use the unigram item probabilities. We employ random hash buckets to avoid evaluating randomly initialized ID embeddings \cite{kang2021learning}. For augmented training, we used an auxiliary Bayesian Personalized Ranking (BPR) loss on the augmentation data \cite{wang2024largelanguagemodelsdata}, which is formulated as follows:
\begin{equation}
L_{aug} = -\sum_{u, i_{pos}, i_{neg}} \log(\sigma(\hat{y}_{u, i_{pos}} - \hat{y}_{u, i_{neg}}))
\end{equation}
Here, $i_{pos}$ and $i_{neg}$ are the cold-start item pair, where the LLM chose $i_{pos}$ over $i_{neg}$ given the full history of user $u$. This auxiliary loss backpropagates gradient signals to cold-start items.
\section{Experiments}
\label{sec:experiments}
To verify the effectiveness of our method, we conducted numerous experiments to address the following research questions:
\begin{itemize}
\item \textbf{RQ1} Does augmentation using the LLM as a user produce better cold-start item performance than the LLM interaction histories?
\item \textbf{RQ2} Which users most improve cold-start item performance with data augmentation?
\item \textbf{RQ3} Do the selected users have better cold-start item performance than unselected users?
\item \textbf{RQ4} Does a learned policy outperform feature‑based user selection for data augmentation in cold‑start item performance?
\end{itemize}
\subsection{Experimental Setup}
We used the Sports and Beauty categories from the public Amazon Product Reviews datasets \cite{He_2016} to evaluate our method. Specifically, we use the 5-core datasets, which iteratively removed users and items with fewer than 5 interactions. Also, we remove items without titles, because we provide item titles in the LLM prompt. Our dataset statistics are described in Table \ref{dataset-table}. Following \cite{wang2024largelanguagemodelsdata}, we used a single-time-point split with a 7:3 ratio to avoid temporal data leakage between the train and test sets. The LLM that we used to generate pairwise preferences is the LLaMa-3.2-3B-Instruct model 
\cite{grattafiori2024llama3herdmodels}. For all experiments before Section \ref{sec:expanding-policy-features}, the policy uses a 5-parameter linear layer followed by Sigmoid. 

For our evaluation, we retrieve the top-K items from all items to measure recall@k, which counts the fraction of test examples where the ground-truth item appears in the top-k retrieved items. Specifically, we evaluate cold (and warm) recall@k \cite{wang2024largelanguagemodelsdata} as defined in Section \ref{sec:reward_generation}. We use the best cold recall@50 after 30 epochs of two-tower training for policy rewards. All experiments after Section \ref{sec:llm-as-user} select 20\% of the warm users for augmentation.

\begin{table}[htbp]
\caption{Amazon Beauty and Sports Dataset Statistics}
\begin{tabular}{||c c c||} 
 \hline
 Statistic & Amazon Beauty &  Amazon Sports\\ [0.5ex] 
 \hline
 \# Users & 22,363 & 35,597 \\ 
 \hline
 \# Items & 12,094 & 18,267 \\
 \hline
 \# Interactions & 198,371 & 295,091 \\
  \hline
\end{tabular}
\label{dataset-table}
\end{table}

\subsection{LLM Effectiveness as a User}
\label{sec:llm-as-user}
To address RQ1, we compare our prompt method using LLM as a user with prior work using the LLM interaction history \cite{wang2024largelanguagemodelsdata}. Using the Amazon Beauty dataset, we evaluate the interaction history method by randomly sampling interactions for augmentation and forming the LLM prompt using the interaction histories. To compare, we evaluate our method using 20\%, 50\% and 100\% of the prior method's data. Table \ref{tab:llm_user_vs_partial_history} illustrates that our method produces higher cold recalls using only 20\% of the interaction history method's augmentation data, proving that the LLM as a user method is better than LLM interaction histories for cold-start item augmentation. 

\begin{table}[h]
\centering
\caption{LLM as a User vs LLM Interaction Cold Recalls (\%)}
\begin{tabular}{|c|c|c|c|} 
\hline
 Method & CR@5 & CR@10 & CR@50 \\
 \hline
 LLM Interaction& $0.09 \pm {0.011}$ & $0.18 \pm {0.013}$  & $1.66 \pm {0.056}$ \\ 
 \hline
 20\% LLM Users & $0.17 \pm {0.0070}$ & $0.41 \pm {0.020}$ & $2.18 \pm {0.065}$  \\
 \hline
 50\% LLM Users & $0.24 \pm {0.022}$ & $0.57 \pm {0.039}$ & $3.54 \pm {0.119}$\\
  \hline
  100\% LLM Users & $0.24 \pm {0.022}$ & $0.62 \pm {0.031}$ & $3.34 \pm {0.089}$\\
  \hline
\end{tabular}
\label{tab:llm_user_vs_partial_history}
\end{table}

Increasing the augmentation data from 20\% to 50\% results in substantial cold recall@5, 10, and 50 increases. However, augmenting 100\% of the users does not lead to any further improvements.

\subsection{Best Users for Augmentation}
\label{sec:best-users}
To address RQ2, we evaluate cold recall: (1) without augmentation, (2) randomly selecting 20\% of users for augmentation, and (3) selecting the top 20\% of users for augmentation per user feature. For each user selection experiment, we report the means and standard errors across 20 two-tower jobs in Table \ref{tab:selection-recalls}. The standard error is computed as $SE = \frac{\sigma}{\sqrt{n}}$ where $\sigma$ and $n$ are the standard deviation and number of two-tower jobs. 

For each dataset, we picked the user features that yielded the five highest average cold recall@50 on the test dataset to be policy inputs. For Amazon Beauty, these features are MP, AP, CSD, AR, and V. The top five Amazon Sports features are RV, CKLD, V, AP, and EE. Surprisingly, these results indicate that users who interacted with popular items (MP, AP) in the Beauty dataset were better augmentation candidates than users who interacted with diverse item categories (CSD) or items before they became trendy (V). This could indicate that the Beauty users' preferences shifted from training to test time: users who were previously interested in popular items later became more interested in cold-start items. Also, selecting users with the highest MP and AP scores caused augmented training to indirectly align cold-start item ID embeddings with the popular items' ID embeddings. We hypothesize that popular item IDs are a good proxy for cold-start item IDs, because they are more aligned with the user IDs. 

For Amazon Sports, the results show that selecting users who interacted with diverse items (CKLD, EE), items before they became popular (V), or rated items with high variation (RV) were favorable candidates for augmentation. Users with high embedding entropy (EE) and category diversity (CKLD) tend to interact with diverse items and exhibit exploratory behavior. They may also rate items more sporadically than other users, because they explore new brands or categories of items which they either strongly like or dislike. Further, users with high CKLD values have category distributions that diverge from the global norm. These types of users are considered "grey sheep" users \cite{10.1145/2930238.2930242}, who have unusual preferences and may be more inclined to review new items. 
Also, users who reviewed items before they became popular (V) may be fashionable and actively seek new items in the catalog. Overall, our findings show that generating cold-start item augmentation examples using specific types of users improves cold recall over randomly selecting users.

\subsection{Selected vs. Unselected Users}
\begin{figure}[]
    \begin{subfigure}[b]{\linewidth}
        \hspace*{-2em}
        \centering
        \includegraphics[width=0.95\linewidth]{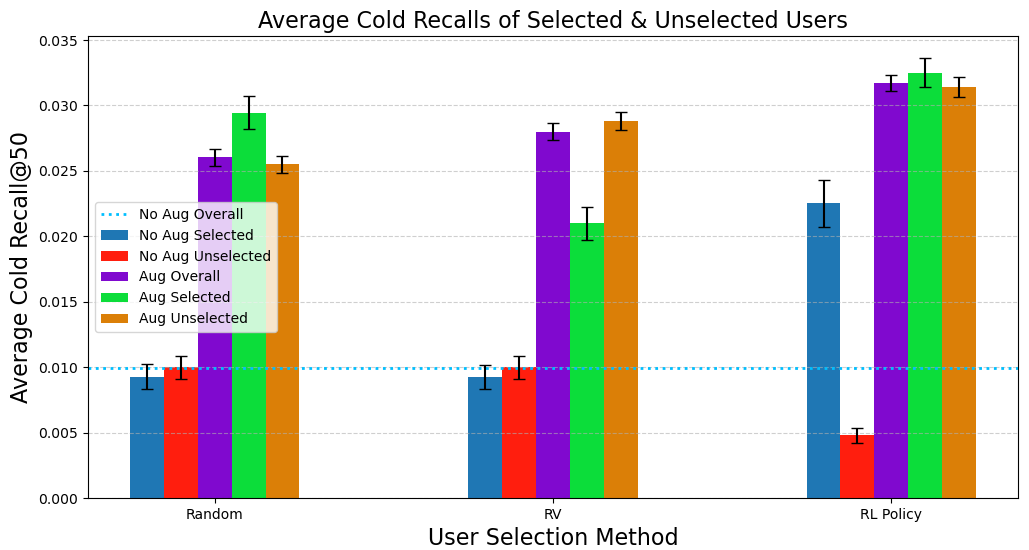}
        \caption{Sports Dataset Bar Chart}
        \label{fig:sports_bar_chart}
    \end{subfigure}
    \begin{subfigure}[h]{\linewidth}
        \hspace*{-2em}
        \centering
        \includegraphics[width=0.95\linewidth]{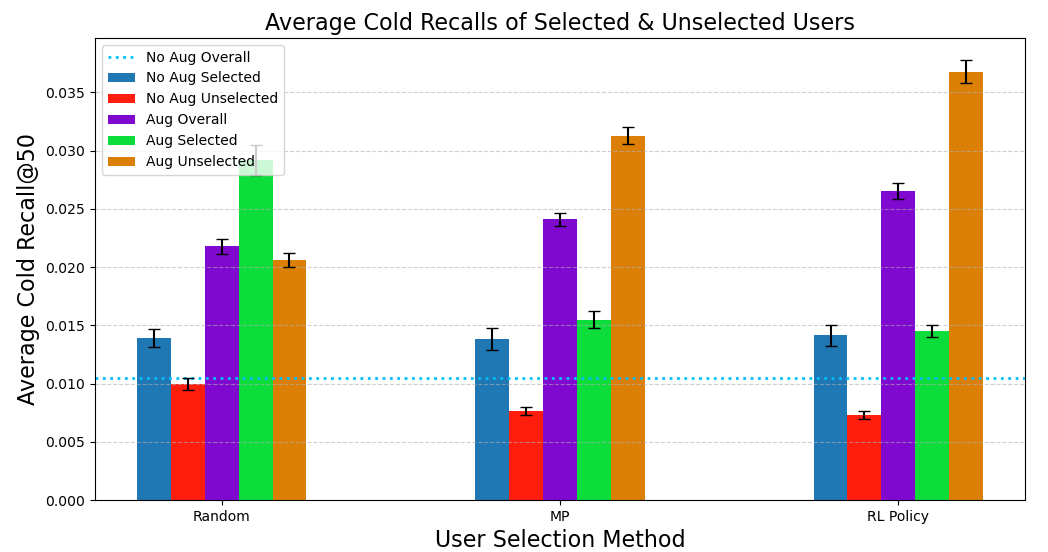}
        \caption{Beauty Dataset Bar Chart}
        \label{fig:beauty_bar_chart}
    \end{subfigure}
\caption{Selected and unselected users' cold recall@50 using three user selection strategies.}
\label{fig:bar_charts}
\end{figure}

To address RQ3, we selected users for augmentation using the same three strategies mentioned at the start of Section \ref{sec:best-users}. For each selection method, we partition the test set into examples where the user was selected and unselected for augmentation. Then, we evaluated cold recall@50 with and without augmentation on both partitions. Across all user selection strategies, we trained the same 20 two-tower models without augmentation. The results for random, best feature, and RL policy user selection on Beauty and Sports are visualized in Figure \ref{fig:bar_charts}. We call the user feature whose top-20\% users yielded the highest cold recall@50 the best feature. 
\begin{table*}[!t]
    \centering
     \caption{Average Cold Recall@50,10 and Warm Recall@50,10 (\%) with and without selecting users for augmentation. The largest and second largest recalls in each column are highlighted in bold and underlined. Augmentation presents a minor warm recall tradeoff for significant cold recall increase. Refer to Table \ref{tab:user-feature-definitions} for the user feature names and definitions.}
    \begin{tabular}{p{2.4cm}|cc|cc|cc|cc}
        \toprule
        & \multicolumn{2}{c|}{Cold Recall@50} & \multicolumn{2}{c|}{Cold Recall@10} & \multicolumn{2}{c|}{Warm Recall@50} & \multicolumn{2}{c}{Warm Recall@10} \\
         Selection Method & Beauty & Sports   & Beauty & Sports    & Beauty & Sports     & Beauty & Sports \\
         \midrule
         None & 1.05 ± 0.05 & 1.15 ± 0.08 & 0.07 ± 0.00 & 0.12 ± 0.01 & \textbf{5.55} ± 0.02 & \textbf{5.53} ± 0.02 & \textbf{1.66} ± 0.02 & \textbf{1.70} ± 0.01 \\
          \hline
         Random & 2.18 ± 0.06 & 2.60 ± 0.07 &  0.41 ± 0.02 & 0.56 ± 0.03 & 4.89 ± 0.04 & 5.13 ± 0.02 &  1.42 ± 0.02 & 1.54 ± 0.01\\
          \hline
         MP & \underline{2.41} ± 0.06 & 2.44 ± 0.07 & \underline{0.50} ± 0.04 & 0.48 ± 0.02 & 4.74 ± 0.03 & 5.00 ± 0.01 & 1.32 ± 0.01 & 1.48 ± 0.01 \\
         AR & 2.19 ± 0.08 & 1.68 ± 0.04 & 0.38 ± 0.04 & 0.29 ± 0.02 & 4.86 ± 0.04 & 5.10 ± 0.02 & 1.41 ± 0.01 & 1.53 ± 0.01 \\
         AP & 2.39 ± 0.06 & 2.58 ± 0.04 & 0.48 ± 0.03 & 0.61 ± 0.03 & 4.71 ± 0.03 & 4.97 ± 0.02 & 1.31 ± 0.01 & 1.53 ± 0.01 \\
         RV & 1.97 ± 0.07 & \underline{2.80} ± 0.07 & 0.26 ± 0.02 & 0.65 ± 0.02 & 4.88 ± 0.04 & 5.03 ± 0.02 & \underline{1.45} ± 0.02 & 1.51 ± 0.01 \\
         CKLD & 1.70 ± 0.05 & 2.78 ± 0.04 & 0.24 ± 0.02 & \underline{0.67} ± 0.02 & \underline{4.92} ± 0.02 & \underline{5.23} ± 0.02 & 1.42 ± 0.01 & \underline{1.60} ± 0.01 \\
         CSD & 2.26 ± 0.07 & 2.50 ± 0.04 & 0.31 ± 0.02 & 0.54 ± 0.02 & 4.79 ± 0.04 & 4.98 ± 0.02 & 1.38 ± 0.01 & 1.48 ± 0.01 \\
         EE & 1.98 ± 0.05 & 2.53 ± 0.06 & 0.39 ± 0.03 & 0.52 ± 0.04 & 4.78 ± 0.03 & 5.00 ± 0.02 & 1.36 ± 0.02 & 1.52 ± 0.01 \\
         V & 2.02 ± 0.07 & 2.63 ± 0.06 & 0.35 ± 0.03 & 0.50 ± 0.02 & 4.73 ± 0.03 & 5.02 ± 0.02 & 1.36 ± 0.01 & 1.54 ± 0.01 \\
         \hline
         Learned Policy & \textbf{2.65} ± 0.06 & \textbf{3.17} ± 0.06 & \textbf{0.55} ± 0.03 & \textbf{0.72} ± 0.03 & 4.69 ± 0.03 & 5.03 ± 0.02 & 1.32 ± 0.01 & 1.54 ± 0.01 \\
          \bottomrule
    \end{tabular}
    \label{tab:selection-recalls}
\end{table*}
Since the results are averaged over 20 jobs, we show standard error bars for each stratified average cold recall@50. Without augmentation, the users selected by the learned policy significantly outperformed the unselected users. 
As shown in Table \ref{tab:cold-recall-increase} and Figure \ref{fig:bar_charts}, augmentation improved the cold recall@50 on both selected and unselected user test splits as expected. Interestingly, for the best feature and learned policy selection methods, the increase in unselected users' cold recall from no augmentation to cold-start item augmentation is larger than the selected users' improvement. This is because each selected user's ID embedding is trained to align with randomly sampled cold-start item ID embeddings from the auxiliary BPR loss. Most likely, they are not the selected users' ground truth items. However, unselected users' ID embeddings are not directly pulled towards any cold-start item ID embeddings. At the same time, the cold-start item ID embeddings are trained to be closer to the general user ID embedding space. This can incidentally cause unselected users' cold recall improvement to exceed that of selected users. For evaluations on the selected and unselected users across all user features, refer to Figures \ref{fig:full_beauty_bar_chart} and \ref{fig:full_sports_bar_chart} in the Appendix.

\begin{table}[htbp]
  \centering
  \caption{Cold recall@50 improvements (\%) with data augmentation on selected and unselected user splits}
  \label{tab:placeholder}
  \begin{tabular}{|l|cc|cc|}
    \hline
        & \multicolumn{2}{p{1.8cm}|}{Selected Users} & \multicolumn{2}{p{2.2cm}|}{Unselected Users}\\
     Selection Method &  Beauty & Sports & Beauty & Sports \\
     \hline 
     Random & 109.92 & 217.46 & 106.83 & 154.75\\
     \hline 
     Best Feature & 12.16 & 126.78 & 310.76 & 188.58  \\
     \hline 
     Learned Policy & 2.69 & 44.38 & 403.15 & 554.17 \\ 
     \hline
  \end{tabular}
  \label{tab:cold-recall-increase}
\end{table}

\subsection{Optimizing Cold Recalls}
To address RQ4, we provided the five user features whose top-k user selection produced the best average cold recall@50 to the policy model. Table \ref{tab:selection-recalls} shows that selecting users using the learned policy lead to higher cold recalls than the best features. The observed improvements are statistically significant, with the right-tailed p-tests for Beauty and Sports yielding: $p_{Beauty}=9.35*10^{-6}$ and $p_{sports}=1.70*10^{-8}$. This test shows that training a policy with user features to optimize for cold recall outperforms top 20\% user selection on individual features. We reported the percentage improvements over the best feature selection baselines in Table \ref{tab:recall-improvements}. For information on policy training, refer to Appendix \ref{sec:policy-training}.
  
\begin{figure}
    \begin{subfigure}{\linewidth}
        \hspace*{-2em}
        \centering
        \includegraphics[width=0.9\linewidth]{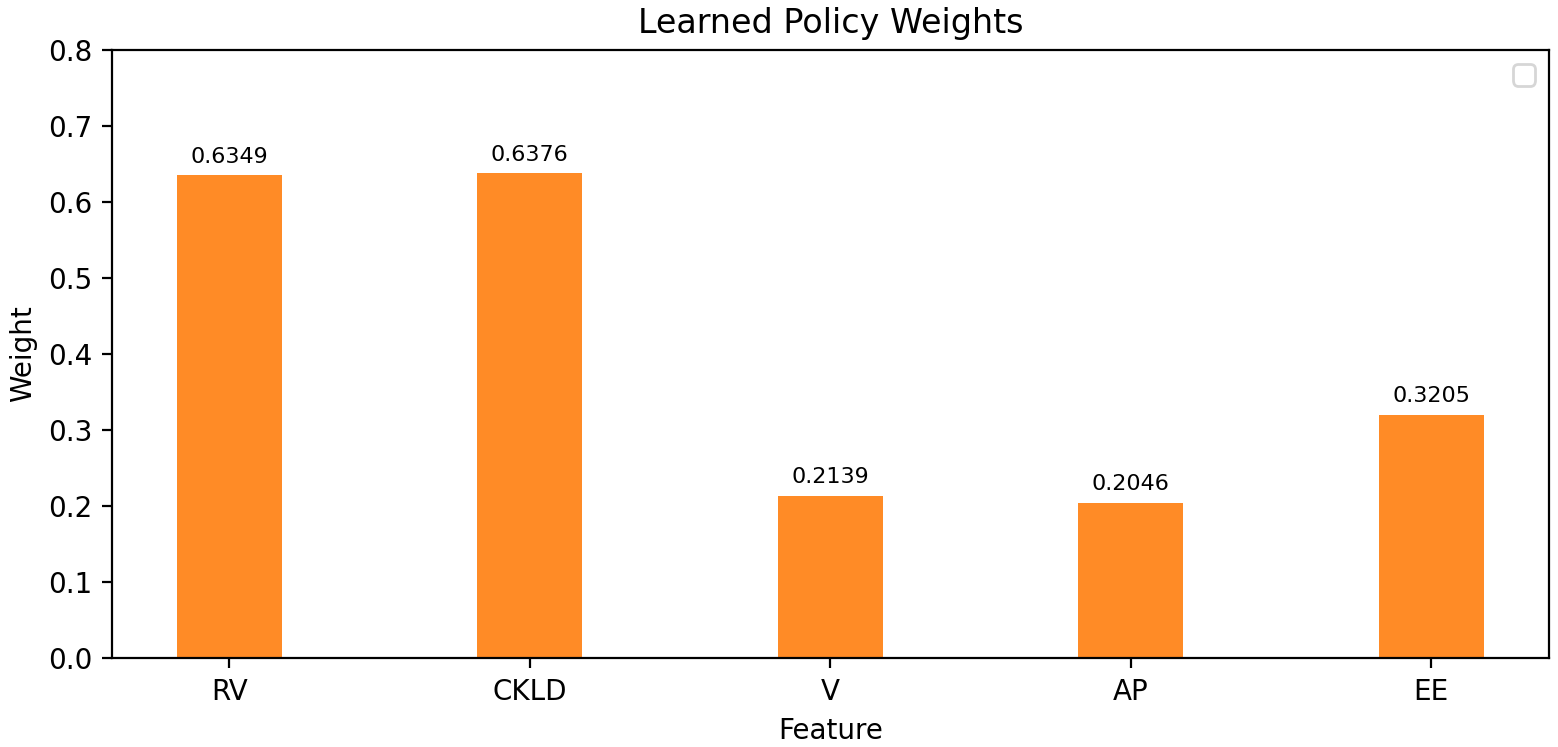}
        \caption{Sports Dataset Policy Weights}
        \label{fig:sports_policy_weights}
    \end{subfigure}
    \begin{subfigure}{\linewidth}
        \centering
        \includegraphics[width=0.9\linewidth]{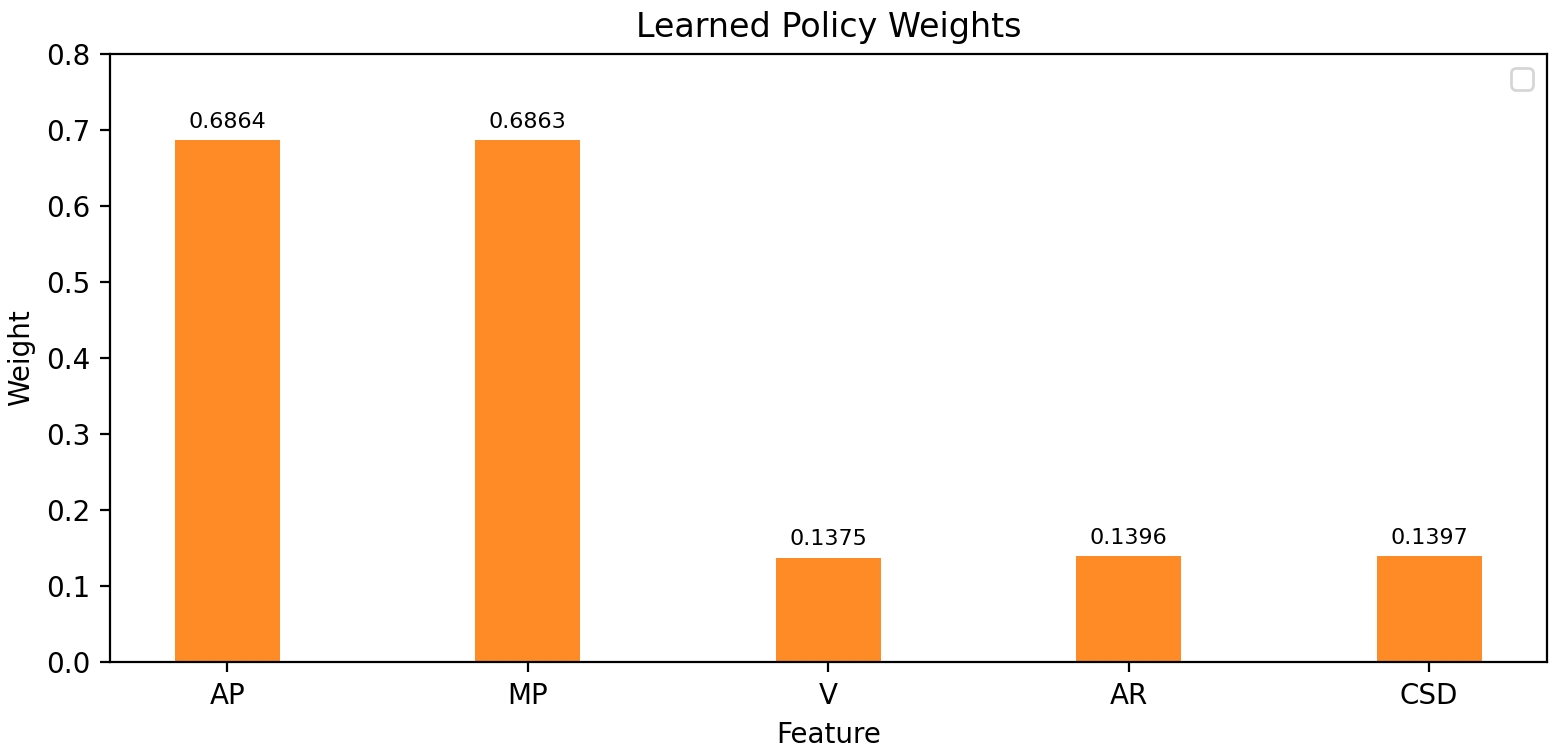}
        \caption{Beauty Dataset Policy Weights}
        \label{fig:beauty_policy_weights}
    \end{subfigure}
\caption{Learned policy weights for each input feature. Refer to Table \ref{tab:user-feature-definitions} for the user feature names and definitions.}
\label{fig:policy_weights}
\end{figure}

\subsection{Comparing Policy Weights}
Before training the policy, we initialize the policy weights of the two best features to higher values than the remaining features. This bootstrapped initialization ensures that the policy initially selects reasonable users for augmentation, then locally optimizes this selection during training. 

From the policy weights displayed in Figure \ref{fig:policy_weights}, the policy preserved emphasis on the stronger features while differentiating among the weaker features. The policy learns to assign higher weights for the features contributing more towards cold recall. On the Sports dataset, the policy significantly increased the weight for the embedding entropy (EE) feature, which yielded a lower cold recall@50 than V and AP in Table \ref{tab:selection-recalls}. This suggests that the policy learns a combination of features that increases cold recall which may not correlate with the individual features. EE is the diversity of a user's historical item metadata embeddings, including item brands, categories, titles, descriptions, and features. Thus, users who interact with unique item categories (CKLD), brands, and titles are good candidates for cold-start item augmentation.

\begin{table}[htbp]
  \centering
  \caption{Cold recall@50 improvements over random and best feature user selection.}
  \label{tab:recall-improvements}
  \begin{tabular}{|l|cc|cc|}
    \hline
    & \multicolumn{2}{p{2.5cm}|}{Improvement over Random Selection} & \multicolumn{2}{p{2.5cm}|}{Improvement over Best Feature} \\
    \hline 
    Method & Beauty & Sports & Beauty & Sports \\
    \hline
    Best Feature & 10.56\% & 7.69\% & -- & -- \\
    Learned Policy    & 21.56\% & 21.92\% & 9.96\% & 13.21\% \\
    \hline
  \end{tabular}
\end{table}
\subsection{Expanding Policy Features}
\label{sec:expanding-policy-features}
To scale up the policy model, we save user embeddings trained via two-tower using PCA \cite{takeshita2025randomly, raunak-etal-2019-effective} and pass them into the policy model along with the original five user features. We use the embeddings from the top layer of the user tower, because it captures more semantic information than the user ID embeddings in the embedding table. We also improve the policy's ability to learn nonlinear relationships between input features by leveraging a two-layer multi-layer-perceptron (MLP). In the forward pass, we first concatenate the original five user features with the compressed user embeddings. Then, the features are passed through Linear, LayerNorm, ReLU, Linear and Sigmoid layers respectively.  

Since training two-tower models for 30 epochs slowly generates rewards, we train two-layer policies using proxy rewards. Our first proxy reward method pretrains two-tower models on IBS. Then, for each iteration of policy training, we fine-tune the pretrained models with the combined IBS and BPR loss for 3 epochs. We also generate proxy rewards by training two-tower models from scratch on the combined loss but stop training after 5 epochs. We compute proxy rewards using the two-tower epochs with the highest cold recall@50. To evaluate two-layer policies trained on proxy rewards, we select the top 20\% of users using the policy weights from the iteration with the highest proxy reward. Then, we train two-tower models for 30 epochs using the selected users for data augmentation. 
\begin{table}[htbp]
  \centering
  \caption{Cold recall@50 (\%) from proxy reward policy training on Amazon Beauty.}
  \label{tab:proxy-reward}
  \begin{tabular}{|l|cc|}
    \hline
    & \multicolumn{2}{|c|}{Proxy Reward} \\
    \hline 
    Policy Features & Fine Tuning & Early Stopping \\
    \hline
    Original Five User Features & 2.49 ± 0.06 & 1.65 ± 0.04 \\
    Additional User Embeddings  & 3.08 ± 0.08 & 2.14 ± 0.07 \\
    \hline
  \end{tabular}
\end{table}
Table \ref{tab:proxy-reward} shows that passing compressed user embeddings and the original five features to the policy improves its performance over passing only the original five features to the policy. Further, fine-tuning the two-tower models serves as a better proxy reward than early stopped models. This is because the fine-tuned models were pretrained on IBS, the primary recommendation task, which allows them to retain knowledge that aligns their rewards more closely with the true rewards than early-stopped two-tower models. 
\section{Conclusion and Discussions}
\label{sec:conclusion}
Recommending cold-start items is challenging, because they don't have any interactions. Existing works use LLM-based data augmentation to address this issue by generating cold-start item interactions for random users. Instead, we frame user selection as a RL problem and train a policy to select users for cold-start item data augmentation. Our experiments avoided user protected attributes (e.g. age, gender, etc) due to ethical concerns. We conducted our experiments on two real-world datasets, demonstrating that a learned policy enhances cold-start item performance substantially. 

Our evaluation only considered single-domain datasets. We suggest that future works run experiments on cross-domain datasets. This could require the policy to use the cold-start item embedding pairs corresponding to each selected user. Then, the policy can leverage the cold-start items' domains to select optimal augmentation examples. Another future work could entail Actor-Critic \cite{mnih2016asynchronous} or PPO \cite{schulman2017proximal} policy implementations to improve stabilization.
\clearpage

\printbibliography 
\section{Appendix}
\subsection{Rationale Behind User Features}
\label{sec:user-features-rationale}
We handcrafted user features to capture a wide variety of user characteristics and enhance cold-start item performance via augmentation. The additional user features we did not include in the policy are:
\begin{enumerate}
    \item Number of Reviews (NR) - Number of items a user reviewed
    \item Median Rating (MR) - Median rating of a user's reviews
    \item Brand Simpson Diversity (BSD) - Sum of squares of a user’s historical item brand probabilities
    \item Brand KL Diversity (BKLD) -  KL-divergence between the user’s historical and global item brands
\end{enumerate}The NR and item popularity features capture users with well trained user or item embeddings, while item rating features (AR, MR, RV) captures satisfied, dissatisfied, or uncertain users. 

\subsection{Reproducibility Details}
Our experiments used 4 NVIDIA RTX 6000 Ada GPUs and 24 AMD EPYC 7713 CPUs. We used Adagrad to train two-tower models, and set the learning rate to $7e-4$, Softmax temperature to $0.1$, cosine similarity to $True$, and dropout rate to $0.01$. The embedding, hidden, and output dimensions for Beauty and Sports are $1024$, $2048$, and $512$, respectively. We use batch sizes $256$ and $512$ for Beauty and Sports, respectively. For the BPR loss, we use a coefficient $0.01$ and augmentation batch sizes $8$ and $16$ for Beauty and Sports, respectively. We train the policy using SGD and set the learning rates to $0.001$ and $0.02$, Sigmoid temperatures to $0.2$ and $0.12$, and initial baseline EMAs coefficients to $0.0$ and $0.9$ for Beauty and Sports, respectively. Both datasets use an EMA coefficient $0.3$ during training. We used Sigmoid temperature annealing for Beauty with a decay of $0.9$ and final temperature of $0.07$. Policy training on true rewards using 24 two-towers per iteration requires approximately 24 GPU hours for 40 iterations. On proxy rewards, using 6 two-tower jobs per iteration requires approximately 2 GPU hours for 100 iterations.

\subsection{Policy Reward Plots}
\label{sec:policy-training}
We trained the five-parameter policies using 24 and 16 two-tower models per iteration for Beauty and Sports, respectively. The red and blue bands represent standard error bounds on the average two-tower cold recall@50 metrics.
\begin{figure}[H]
    \centering
    \begin{subfigure}[h]{0.94\linewidth}
        \includegraphics[width=0.94\linewidth]{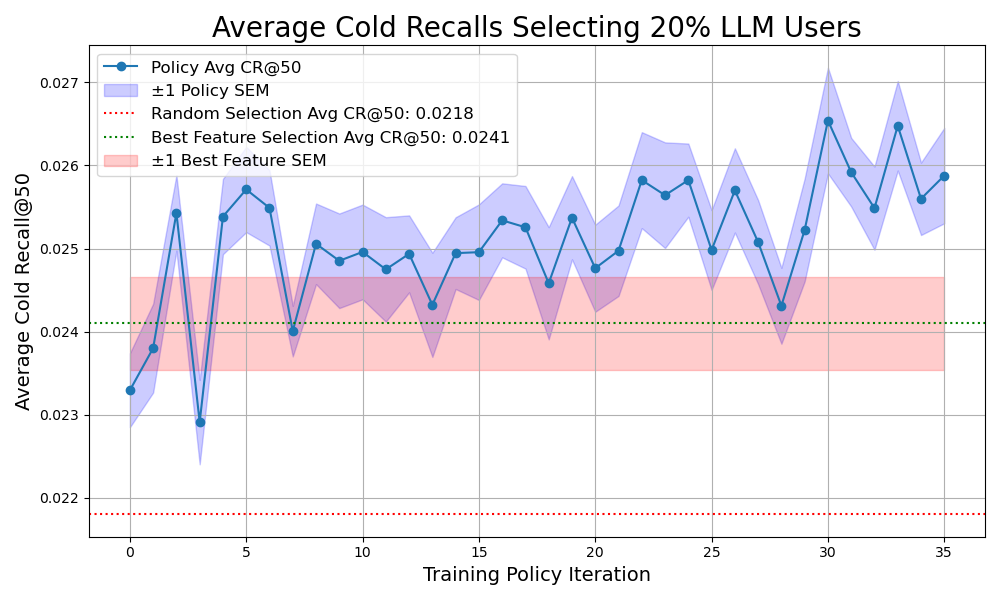}
        \caption{Beauty Policy Training on True Rewards}
        \label{fig:beauty_rl_policy}
    \end{subfigure}
    \begin{subfigure}[h]{0.94\linewidth}
        \includegraphics[width=0.94\linewidth]{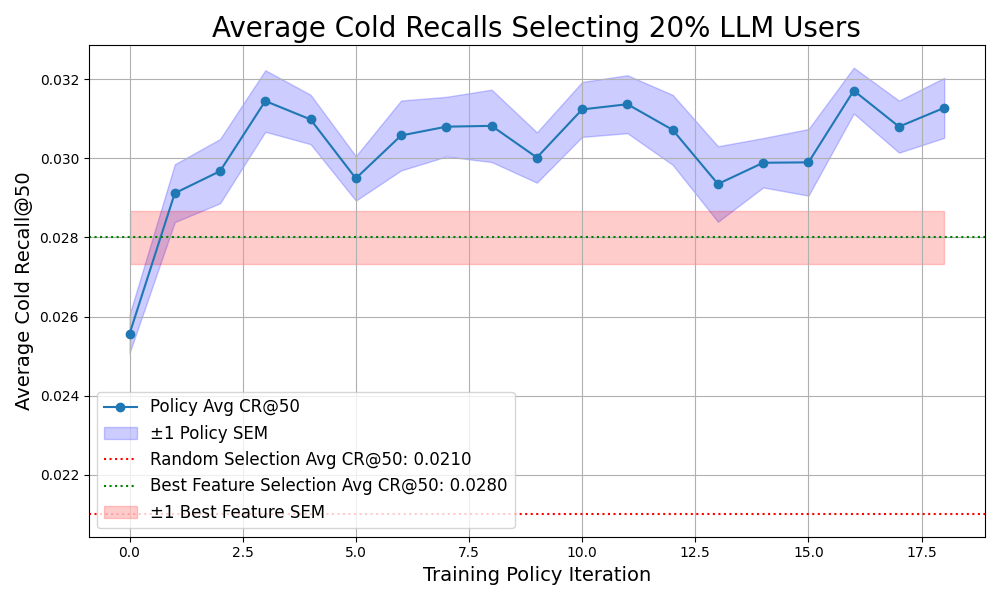}
        \caption{Sports Policy Training on True Rewards}
        \label{fig:sports_rl_policy}
    \end{subfigure}
\end{figure}
\subsubsection{Policy Training on Proxy Reward} Using Amazon Beauty, we compressed the user embeddings to 10 dimensions and trained two-layer policies using both the original and compressed features. The policy has an input layer with 15 features, hidden dimension of 5, and output dimension of 1. Below are the fine-tuned and early-stopped proxy reward plots.
\label{sec:fine-tuned-proxy-reward}
\begin{figure}[H]
    \centering
    \begin{subfigure}[]{\linewidth}
        \includegraphics[width=\linewidth]{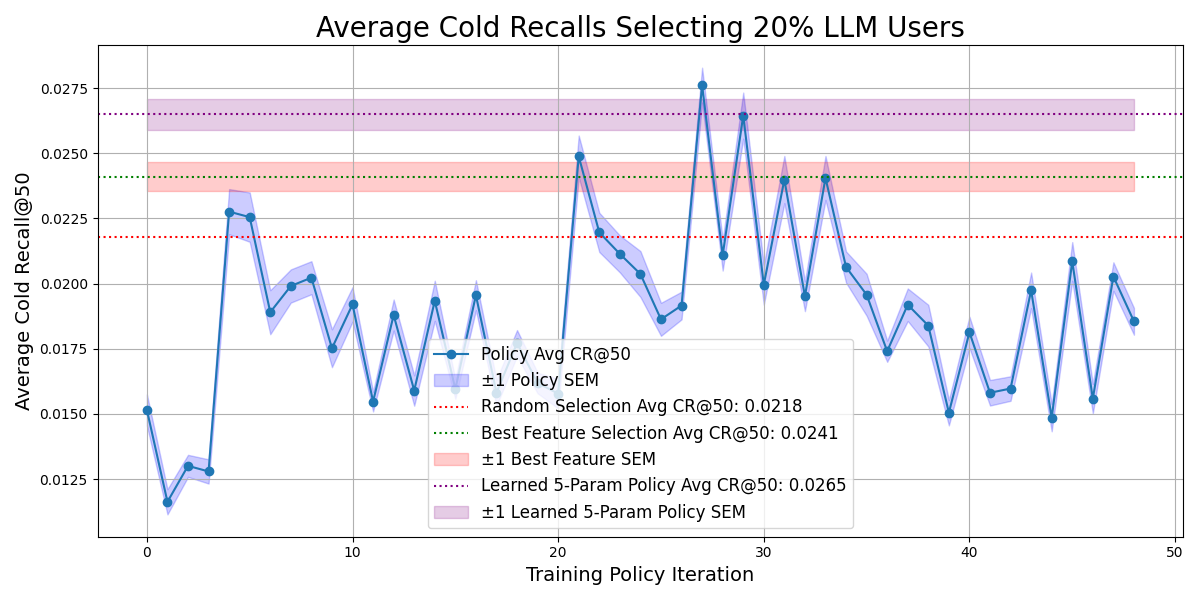}
        \caption{Policy Training on Fine-Tuned Reward Without User IDs}
    \end{subfigure}
    \begin{subfigure}[]{\linewidth}
        \includegraphics[width=\linewidth]{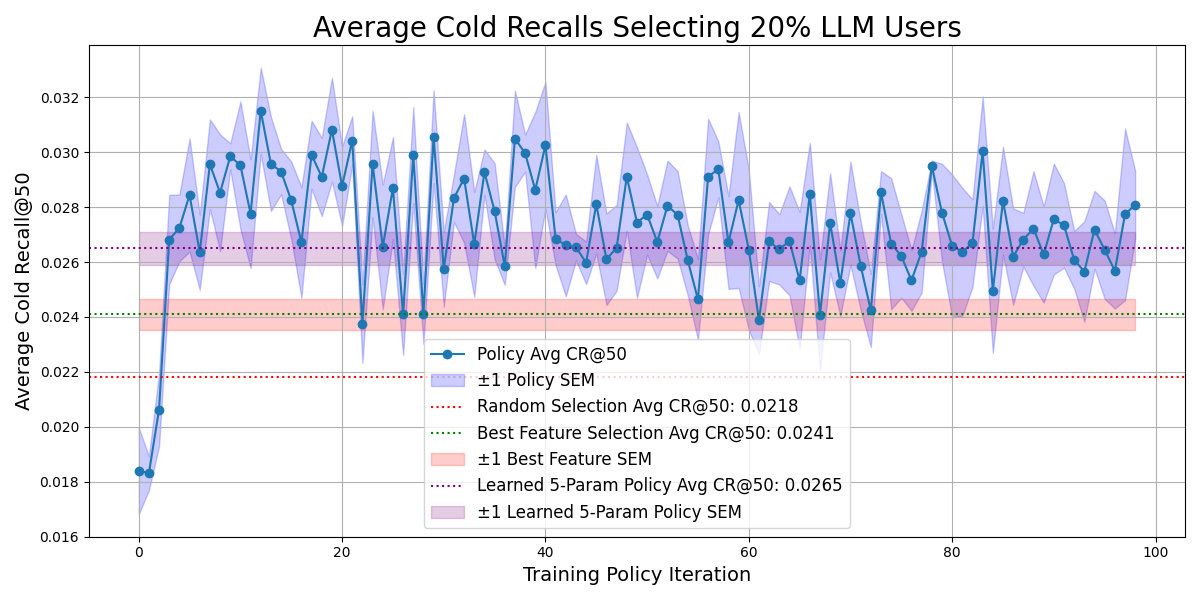}
        \caption{Policy Training on Fine-Tuned Reward With User IDs}
    \end{subfigure}
\end{figure}
\begin{figure}[H]
    \centering
    \begin{subfigure}[h]{\linewidth}
        \includegraphics[width=\linewidth]{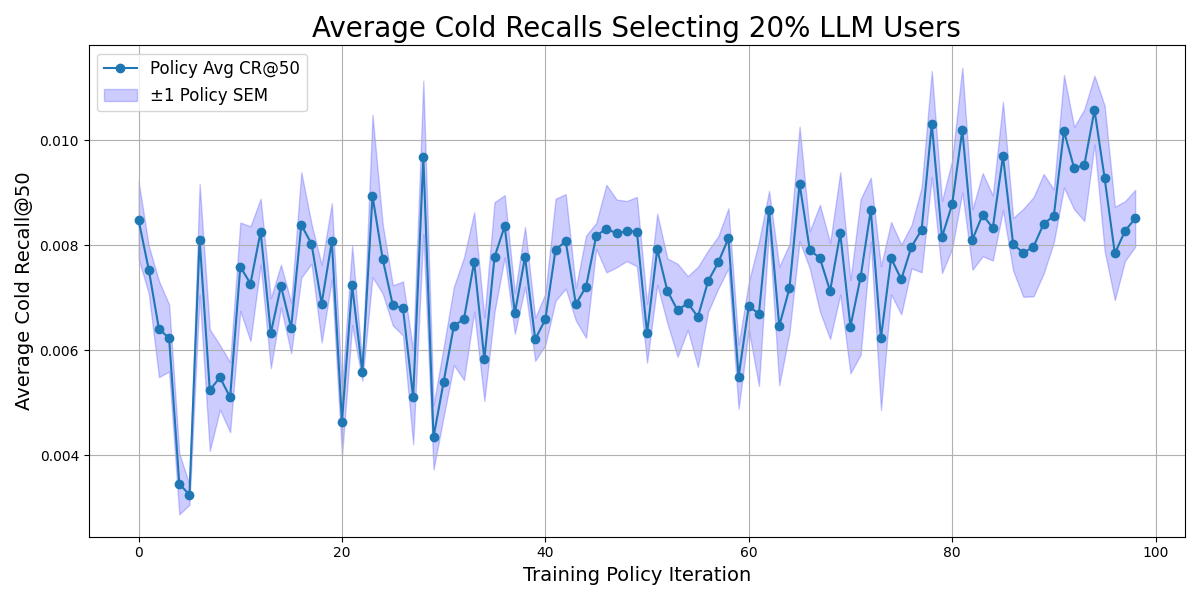}
        \caption{Policy Training on Early-Stopped Reward Without User IDs}
    \end{subfigure}
    \begin{subfigure}[h]{\linewidth}
        \includegraphics[width=\linewidth]{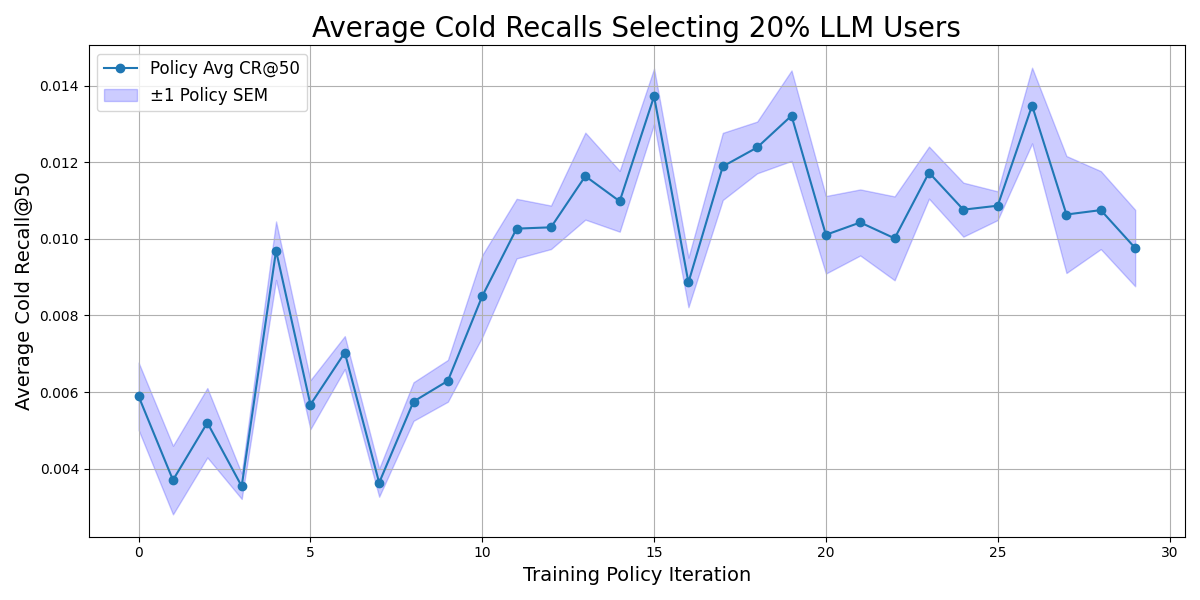}
        \caption{Policy Training on Early-Stopped Reward With User IDs}
        \label{fig:sports_rl_policy}
    \end{subfigure}
\end{figure}
\clearpage

\subsection{Selected and Unselected Users' Bar Charts}
\begin{figure}[H]
    \begin{subfigure}[b]{\textwidth}
        \includegraphics[width=1.02\textwidth]{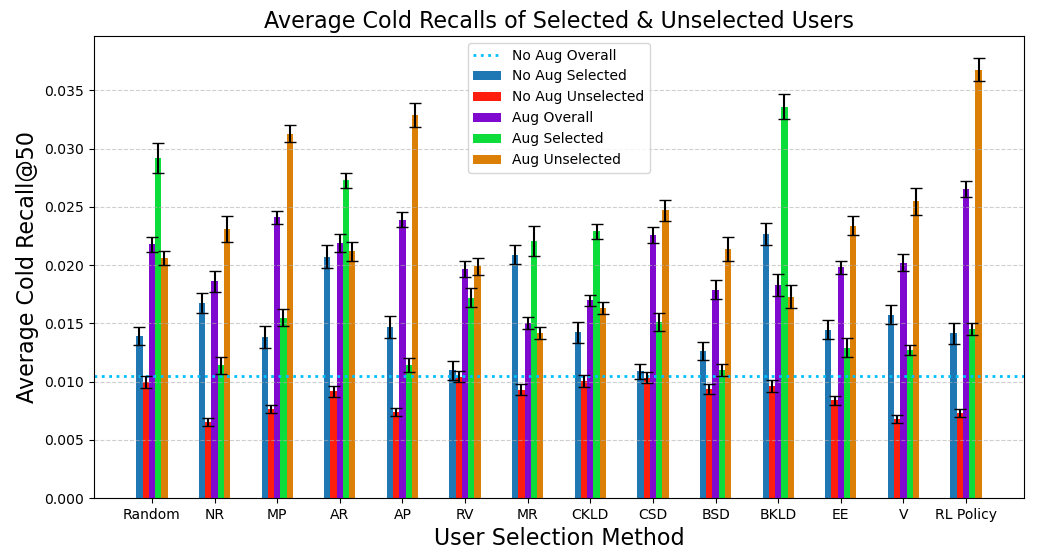}
        \caption{Beauty dataset bar chart across all user features}
        \label{fig:full_beauty_bar_chart}
    \end{subfigure}
    \hfill
    \begin{subfigure}[b]{\textwidth}
       \includegraphics[width=1.02\textwidth]{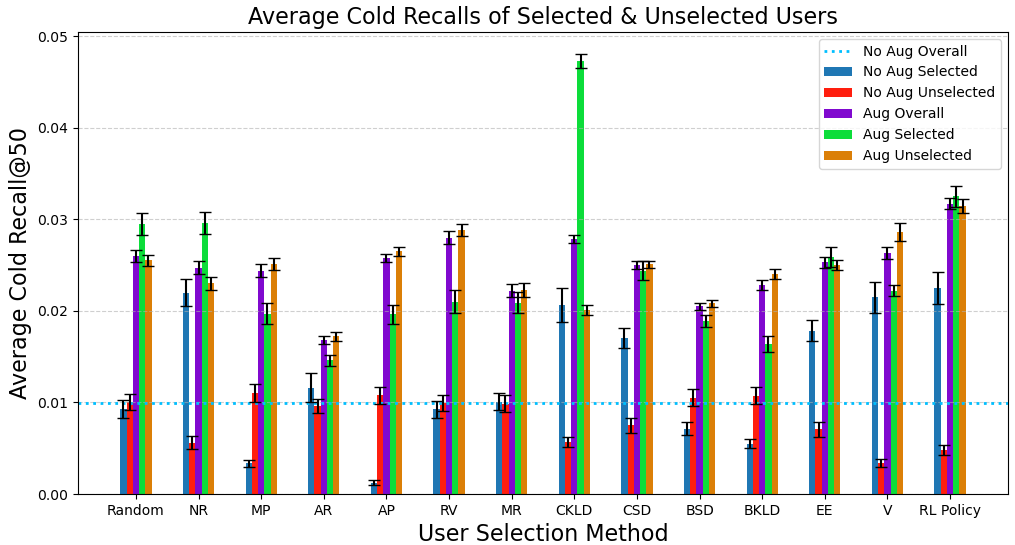}
        \caption{Sports dataset bar chart across all user features}
        \label{fig:full_sports_bar_chart}
    \end{subfigure}
\end{figure}

\end{document}